\documentclass[11pt,dvips]{article}

\usepackage{amssymb,amsfonts}     
 \usepackage{graphicx}
\usepackage{epsfig}
\usepackage{amsmath}
\usepackage{amsbsy}

\usepackage{mathrsfs}

\begin{document}

\title{{\bf {\LARGE Developing de Broglie Wave\footnote{In: Progress in Physics, v.4, 32-35, 2006; recompiled, self-contained treatment of one of topics from: J.X. Zheng-Johansson and P-I. Johansson, {\it Unification of Classical, Quantum and Relativistic Mechanics and of the Four Forces},  Nova Sci. Pub., NY, 2nd printing, later 2006, a revised and enlarged edition of the 1st printing of April 2006. }
}}}
\author{J X  Zheng-Johansson$^1$ and P-I Johansson$^2$  
\\
{\small  {\it 1. Inst. of Fundamental Physics Research (IOFPR),  611 93 Nyk\"oping, Sweden. }}
\\
{\small  {\it  2. Uppsala University, Studsvik, 611 82 Nyk\"oping, Sweden}}
\\
{\small August, 2006}
} 
\date{}
\maketitle

\def\d{\delta}
\def\vel{v}
\def\Lamd{{\mit \Lambda}_d{}}
\def\Pfp{\Pm_\vel}
\def\Efp{\Eng_\vel}
\def\imc{\mbox{\scriptsize{vir}}}
\def\Pm{P}
\def\Dc{a_{\Cssub}}
\def\Cssub{{\mbox{\tiny${C}$}}}

\def\Mch{\mathfrak{M}}
\def\pac{\mathscr{Y}}
\def\beat{{\rm b}}
\def\vphi{\varphi}
\def\a{\alpha}
\def\D{\Delta}
\def\th{\theta}
\def\r{{\mbox{\tiny${R}$}}}
\def\re{{\mbox{\tiny${R}$}}}
\def\Xsub{{\mbox{\tiny${X}$}}}
\def\Xssub{{\mbox{\tiny${X}$}}}

\def\Ksub{{\mbox{\tiny${K}$}}}
\def\W{{{\mit \Omega}}}
\def\Wd{\W_d{}}
\def\Nu{{\cal V}}
\def\Nud{\Nu_d{}}
\def\Eng{{\cal E}}
\def\eng{{\varepsilon}}
\def\Kd{K_d{}}
\def\Lam{{\mit \Lambda}}
\def\lam{\lambda}
\def\dagsup{{\mbox{\tiny${\dagger}$}}}
\def\ddagsup{{\mbox{\tiny${\ddagger}$}}}
\def\w{\omega{}}
\def\wdlow{\omega_d }
\def\g{\gamma{}} 
\def\Phimit{{\mit \Phi}}
\def\Psimit{{\mit \Psi}}
\def\lf{\left}
\def\rt{\right}
\def\Kdsub{{\mbox{\tiny${K_d}$}}}
\def\hquad{ \ \ } 
\def\Taum{{\mit \Gamma}}
\def\Tssub{{\mbox{\tiny${T}$}}}
\def\Kssub{{\mbox{\tiny${K}$}}}
\def\Xssub{{\mbox{\tiny${X}$}}}
\def\Wssub{{\mbox{\tiny${W}$}}}

\begin{abstract}
The electromagnetic component waves, comprising together with their generating oscillatory massless charge a material particle, will be Doppler shifted when the charge hence particle is in motion, with a velocity $v$, as a mere mechanical consequence of the source motion.  We illustrate here that two such component waves generated in opposite directions and propagating at speed  $c$ between  walls in a one-dimensional box, superpose into a traveling beat wave of wavelength ${\mit\Lambda}_d$$=(\frac{v}{c}){\mit\Lambda}$ and phase velocity $c^2/v+v$ which resembles directly L. de Broglie's hypothetic phase wave. This phase wave in terms of transporting the particle mass at the speed $v$ and angular frequency ${\mit\Omega}_d=2\pi v /{\mit\Lambda}_d$, with ${\mit\Lambda}_d$ and ${\mit\Omega}_d$ obeying the de Broglie relations, represents a de Broglie wave. The standing-wave function of the de Broglie (phase) wave and its variables for particle dynamics in small geometries are equivalent to the eigen-state solutions to Schr\"odinger equation of an identical system. 
\end{abstract}

\section{Introduction}
As it stood at the turn of the 20th century, M. Planck's quantum theory suggested that energy ($\eng$) is associated with a certain periodic process of frequency ($\nu$), $\eng=h \nu$; and A. Einstein's mass-energy relation suggested the total energy of a particle ($\eng$) is connected to its mass  ($m$), $\eng=mc^2$. Planck and Einstein together implied that mass was associated with a periodic process $mc^2=h \nu$,  and accordingly a larger $\nu$ with a moving mass. Incited by such a connection but also a clash with this from Einstein's relativity theory which suggested a moving mass is associated with a slowing-down clock and thus a smaller $\nu$, 
  L. de Broglie put forward in  1923 \cite{deBroglie1923} 
a hypothesis that a matter particle (moving at velocity $v$) consists of an internal periodic process describable as a packet of phase waves of frequencies dispersed about $\nu$, having a phase velocity $W=\frac{\nu}{k}
=c^2/v$, with $c$ the speed of light, and a  group velocity 
of the phase-wave packet 
equal to $v$. Despite the hypothetic phase wave appeared supernatural and is today not held a standard physics notion, the de Broglie wave has proven in modern physics 
to depict accurately the matter particles, and the de Broglie relations proven their fundamental relations.
 
So inevitably the puzzles with the de Broglie wave 
persist, involving the hypothetic phase waves or not, and are unanswered prior to our recent unification work\cite{Unif1}: What is waving with a de Broglie wave and more generally Schr\"odinger's wave function? If de Broglie's  phase wave is indeed a reality, what is then transmitted at a speed ($W$) being  $\frac{c}{v}$ times the speed of light $c$? How is the de Broglie (phase) wave related to the particle's charge, which if accelerated generates according to Maxwell's theory electromagnetic (EM) waves of speed $c$, and how is it in turn related to  the EM waves, which are commonplace emitted or absorbed by a particle which changes its internal state? In \cite{Unif1} we showed that a physical model able to yield all of the essential properties of a de Broglie particle, in terms of solutions in a unified framework of the three basic mechanics, is provided by a single harmonic oscillating, massless charge $+e$ or $-e$ (termed a {\it vaculeon}) and the resulting electromagnetic waves. The solutions for a basic material particle generally in motion, with the charge quantity  (accompanied with a spin) and energy of the charge as the sole inputs, predict accurately the inertial mass, total wave function, total energy equal to the mass times $c^2$, total momentum,  kinetic energy and linear momentum of the particle, and that the particle is a de Broglie wave, it obeys Newton's laws of motion, the de Broglie relations, the Schr\"odinger equation in small geometries,  Newton's law of gravitation, and the Galileo-Lorentz-Einstein transformation at high velocities. In this paper we give a self-contained illustration of the process by which the  electromagnetic component waves of such a particle in motion superpose into a de Broglie (phase) wave.

\section{Particle;  component waves; dynamic variables}
A free massless vaculeon charge ($q$) endowed with a kinetic energy $\Eng_q$ at its creation, being not dissipatable
except in a pair annihilation, will tend to move about in the vacuum, and yet at larger displacement restored, fully if  
$\Eng_q$ below a threshold, toward equilibrium by the potential field of the surrounding dielectric vacuum being now polarized under the charge's own field \cite{Unif1}. As a result the charge oscillates in the vacuum, at a frequency $\W_q$; once in addition uni-directionally driven, it will also be traveling  at a velocity $v$ here along $X$-axis in a one-dimensional box of length $L$, firstly in $+X$-direction. Let axis $X'$ be attached to the moving charge, $X'= X-\vel T$; let $v$ be low so that $(v/c)^2\rightarrow 0$, with $c$ the velocity of light; accordingly $T'=T$. 
The charge will according to Maxwell's theory generate electromagnetic waves to both $+X$- and $-X$- directions, having the standard plane wave solution, given in dimensionless displacements (of the medium or fields):
\begin{eqnarray}
\label{eq-ux1p}
\varphi^{\dagsup} (X',T)
= C_1 \sin[K^{\dagsup} X'-\W^{\dagsup}  T +\alpha_0],  \quad ({\rm a})
\cr
      \varphi^{\ddagsup}(X',T)=
         - C_1 \sin[K^{\ddagsup}  X'+\W^{\ddagsup} T -\alpha_0],  \quad({\rm b})
\end{eqnarray}
 where $\left[{K^{\dagsup}  \atop K^{\ddagsup}}\right]
=\lim_{(v/c)^2\rightarrow 0} 
\left[{k^{\dagsup}  \atop k^{\ddagsup}}\right]
=K\pm K_d$, with 
$\left[{k^{\dagsup}  \atop k^{\ddagsup}}\right]= \frac{K}{1\mp v/c}$ wavevectors Doppler-shifted due to the source motion from their zero-$v$ value, $K$;   ${\mit\Lambda}=2\pi/K$. 
On defining $k_d= \sqrt{(k^{\dagsup}-K) (K-k^{\ddagsup}) }=\lf(\frac{v}{c}\rt)k$, with  
$k=\g K$ and $\g=1/\sqrt{1-(v/c)^2}$,  its classic-velocity limit gives,  
\begin{eqnarray}   \label{eq-Kd}
K_d=\lim_{(v/c)^2\rightarrow 0} k_d 
=\lf(\frac{v}{c}\rt)K;
\end{eqnarray}
$\left[{\W^{\dagsup}\atop \W^{\ddagsup}} \right]
=\left[{K^{\dagsup}c \atop K^{\ddagsup}c }\right]=\W \pm K_d c$, with   $\W=cK$; $\W=\W_q$ for the classical electromagnetic radiation; 
and $\a_0$ is the initial phase. 
Assuming $\Eng_q$ is large and radiated in $J(>>1)$ wave periods if without re-fuel, the wavetrain of $\varphi^j$ of a length $L_\varphi=J L$ will wind about the box $L$ in $J>>1$ loops. 

The electromagnetic wave $\varphi^j$ of angular frequency $ \w^j=k^jc$, $j=\dagger $ or $ \ddagger$, has according to M. Planck a wave energy $\eng^j$ $= \hbar \w^j $, with $2\pi\hbar=h$ the Planck constant. The waves are here the components of a particle;   
the geometric mean of their wave energies, $\sqrt{\eng^{\dagsup}\eng^{\ddagsup}}$ $
=\hbar\sqrt{\w^{\dagsup}\w^{\ddagsup}}=\g \hbar \W$
gives thereby the total energy of the particle.  
 $\eng_v=\g \hbar \W -\hbar \W =\frac{\hbar}{2} (\frac{v}{c})^2\W [1+\frac{3}{4}(\frac{v}{c})^2+\ldots]$  gives further the particle's kinetic energy and in a similar fashion 
its linear momentum $p_v$ (see \cite{Unif1}); and 
 \begin{eqnarray}
\label{eq-Engv}
&\Eng_v=\lim_{(v/c)^2 \rightarrow 0}\eng_v=\frac{1}{2}\hbar  \left(\frac{v}{c}\right)^2 \W, 
\\  
\label{eq-Pv}
 & P_v=\lim_{(v/c)^2 \rightarrow 0}p_v=\sqrt{2m_0 \Eng_v}
= \hbar  \left(\frac{v}{c}\right)K.
\end{eqnarray} 
   The above continues to indeed imply as L. de Broglie noted that a moving mass has a larger  $\g \W/2\pi$ ($=\nu$), and thus a clash with the time-dilation of Einstein's moving clock. This conflict however vanishes when the underlying physics  becomes clear-cut[2,2006b].

\section{Propagating total wave of particle
}\label{SecI.V} \label{Sec-NdbTrw} 

A tagged wave front of say $\varphi^{\dagsup} (X',T)$ generated by the vaculeon charge, of $v>0$, to its  right at location $X'$ at time $T$, will after a round-trip of distance $2L$ in time $\delta T=2L/c$  return from left, and propagate again to the right  to $X'$ at time $T^*=T+\delta T$. Here it gains  a total extra phase $\alpha' =K2L +2\pi$ due to $2L$ (with $\frac{K^{\dagsup} +K^{\ddagsup} }{2} =K$) and the twice reflections at the massive walls, and becomes  
$$
\varphi^{\dagsup}_r(X', T^*) 
= C_1 \sin[K^{\dagsup} X'-\W^{\dagsup} T +\alpha_0+\alpha'].  \eqno(\ref{eq-ux1p}{\rm a})'
$$
    $\varphi_r^{\dagsup} $ meets with $\varphi^{\dagsup} (X',T^*)$ just generated to the right, an identical wave except for an $\a'$, and superpose with it to a maximum if assuming $K2L=N2\pi$, $N=0,1, \ldots$, returning the same $\varphi^{\dagsup} $ (assuming normalized). Meanwhile, $\varphi^{\dagsup}_r(X', T^*)$ meets $\varphi^{\ddagsup} (X'_1,T^*)$ just 
\begin{figure}[h]
\vspace{0.2cm}
\begin{center}
\includegraphics[width=0.85\textwidth]{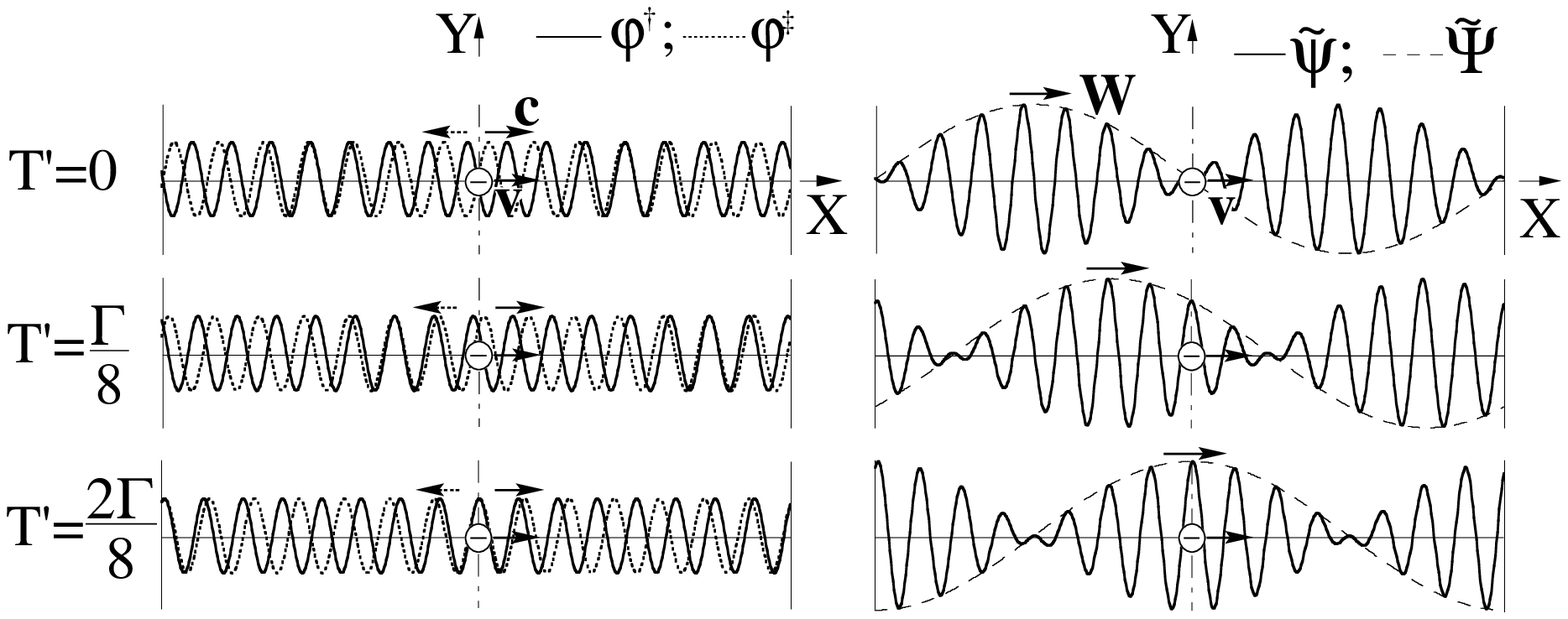}
\end{center}
\vspace{-1.05cm}
\begin{center}
\includegraphics[width=0.85\textwidth]{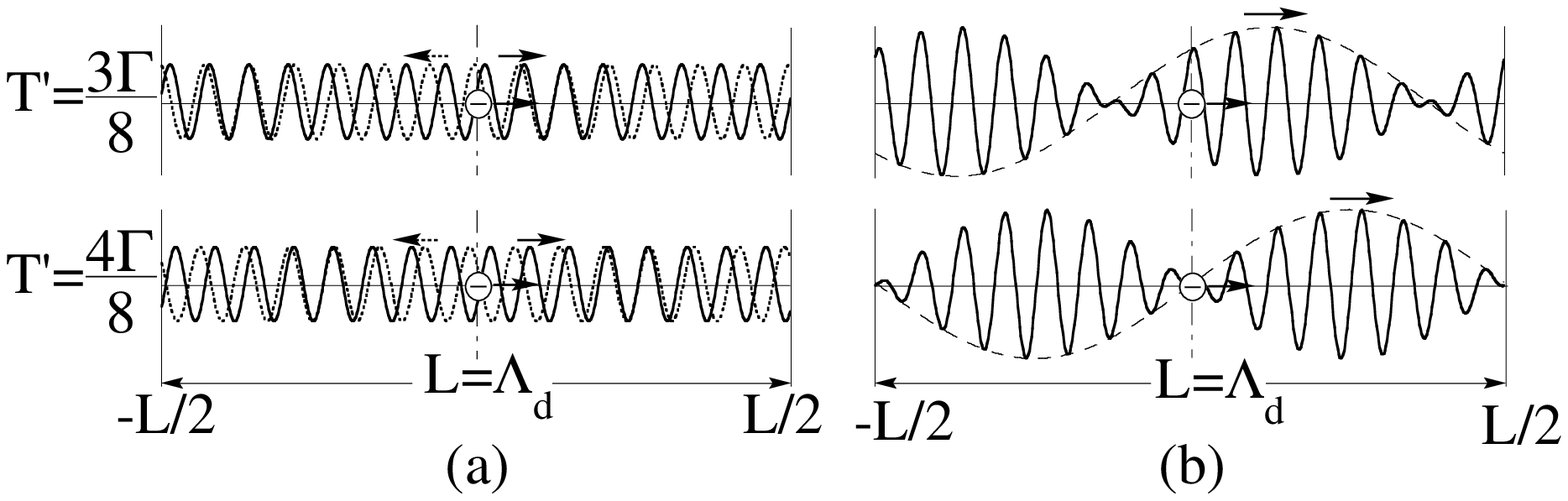}
\end{center}
\vspace{-0.5cm}
 \caption{ 
(a) The time development of electromagnetic waves  with wave speed $c$ and wavelength ${\mit\Lambda}$, $\varphi^{\dagsup}$  generated to the right of (\ref{eq-ux1p}{\rm a})$'$ and $\varphi^{\ddagsup}$ to the left of (\ref{eq-ux1p}{\rm b})  by  a charge  ($\ominus$) traveling   at velocity $\vel$ in $+X$ direction in a one-dimensional box of side $L $.  
(b)  $\varphi^{\dagsup}$ and 
 $\varphi^{\ddagsup}$ superpose to  a beat, or de Broglie phase wave $\widetilde{\psi}$ of (\ref{eq-Ad1}) traveling at phase velocity $W\simeq \frac{c^2}{\vel}$, of wavelength ${\mit\Lambda}_d$.
For the plot: ${\mit\Lambda}=0.067 {\mit\Lambda}_d $, and $\a_0=-\frac{\pi}{2}$; $T'=T-\frac{\Taum}{4}$; $\vel=(\frac{{\mit\Lambda}}{{\mit\Lambda}_d}) c \ll c$.
}
\label{figI.4-dBwav-trv}
\end{figure} 
generated by the charge to the left (Figure \ref{figI.4-dBwav-trv}a), and superpose with it to ${\widetilde \psi}=\varphi^{\dagsup}_r +\varphi^{\ddagsup}$. Using the trigonometric identity (TI), denoting  ${\widetilde \psi}(X',T)={\widetilde \psi}(X',T^*)$,
this is 
$
{\widetilde \psi}(X',T)$ $ =2C_1\cos (KX' -K_d cT )$ $\times \sin(K_d   X'-$ $\W T+\a_0 )
$.
With $X'=X-vT$,  we have  on the $X$-axis:
\begin{eqnarray}
  \label{eq-Ad1}\label{eq-ada}
&&\widetilde{\psi}(X,T) 
=\widetilde{\Phimit}(X,T) \widetilde{\Psimit}(X,T),
              \qquad \qquad \qquad \qquad
\\ 
 \label{eq-Ad1a}
&&\widetilde{\Phimit} (X,T)= 2 C_1 \cos(KX-2K_dcT), \qquad\qquad
\\
 \label{eq-Ad1b}
&&\widetilde{\Psimit}  (X,T)=\sin[\Kd X- (\W+\W_d)T+\a_0], \qquad
\end{eqnarray} 
where $Kv=K_d c$, and 
\begin{eqnarray}\label{eq-nud4}
\W_d = K_d \vel =\lf(\frac{v}{c}\rt)^2\W.
\end{eqnarray} 
$\widetilde{\psi}$ expressed by (\ref{eq-ada}) is 
a {\it traveling beat wave}, as plotted versus $X$ in  Figure \ref{figI.4-dBwav-trv}b  
 for consecutive time points during $\Taum/2$, or Figure \ref{fig-wvpac}a during $\Taum_d/2$. $\widetilde{\psi}$ is due to all the component waves of the charge of the particle while moving in one direction, and thus represents the (propagating) total wave of the particle, to be identified as a {\it de Broglie phase wave} below.

$\widetilde{\psi}$ in (\ref{eq-ada}) has one product component $\widetilde{\Phimit}$ oscillating rapidly on the $X$-axis with the wavelength ${\mit\Lambda}=2\pi/K$, and propagating at the speed of light $c$ at which the total wave energy is transported. 
The other, $\widetilde{\Psimit}$, envelops about $\widetilde{\Phimit}$,  modulating it into  a  slow varying beat $\widetilde{\psi}$.  $\widetilde{\Psimit}$, thus $\widetilde{\psi}$, has a wavevector, wavelength  and angular frequency:
\begin{eqnarray}\label{eq-Wbeat1}
K_{\beat }=K_d, \quad {\mit\Lambda}_{\beat}=\frac{2\pi}{K_{\beat}}=\frac{2\pi}{K_{d}}={\mit\Lambda}_d, \quad
\W_{\beat }= \W +\W_d;
\\ \label{eq-lamd}
{\rm where} \quad {\mit\Lambda}_d=
  \lf(\frac{c}{\vel}\rt){\mit\Lambda}.\qquad\qquad \qquad
\end{eqnarray}
   As follows (\ref{eq-Wbeat1}),  the beat $\widetilde{\psi}$ travels at the {\it phase velocity}  
$$ \refstepcounter{equation} 
\label{eq-V1}
 W= \frac{ \W_{\beat } }{K_{\beat }  } 
=  \frac{\W}{\Kd}  +\vel =  \left(\frac{c}{\vel}\right)c+\vel.
\eqno(\ref{eq-V1})
$$  

\section{De Broglie wave}
The beat wave, of the wave variables ${\mit\Lambda}_d$ and  $ K_d =2\pi/{\mit\Lambda}_d$, transports with it  the  mass of the particle at the velocity $v$, and in this context  
represents thereby a periodic process of (the center of mass of) the particle, of a wavelength and wavevector equal to ${\mit\Lambda}_d$ and $K_d$ of the beat wave. $K_d$ and $v$ define for the particle dynamics an angular frequency, $K_d v =\W_d$, as expressed by (\ref{eq-nud4}).
Combining (\ref{eq-lamd}) and  (\ref{eq-Pv}), and (\ref{eq-nud4}) and (\ref{eq-Engv}) yield just the {\it de Broglie relations}:
 \refstepcounter{equation}\label{eq-px3}
  \refstepcounter{equation}\label{eq-e3}\label{eq-e3b}
$$\displaylines{
\hfill\qquad \Pfp  = \frac{h}{{\mit\Lambda}_d};  \qquad\quad\hfill
(\ref{eq-px3}) \hfill\hfill
\qquad\quad
 \Efp  = \frac{1}{2} \hbar \W_d. \qquad  \hfill\quad(\ref{eq-e3}) \nonumber
}$$ 
   Accordingly $K_d$, ${\mit\Lambda}_d$, and  $\W_d$ represent the de Broglie wavevector, wavelength and angular frequency. The beat wave $\widetilde{\psi}$ of a phase velocity $W$ resembles thereby the {\it de Broglie phase wave} and it  in the context of transporting the particle mass represents the {\it de Broglie wave} of the particle.

\section{Virtual source. Reflected total particle wave}
At an earlier time $T_1=T-\Delta T$,  at a distance $L$ advancing its present location $X$, with $\Delta T =L/v$, the actual charge was traveling to the left, let axis $X''(=X+\vel T)$ be fixed to it. This past-time charge, said being virtual, generated similarly at location $X''$ at time  $T_1$ one component wave  ${\varphi^{\dagsup}}^{\imc}(X'',T_1^*)$ to the right,  which after traversing $2L$ returned from left to $X''$ at time $T_1^* =T_1+\delta T$   as ${\varphi^{\dagsup}}^{\imc}_r(X'',T_1^*)
=C_1 \sin(K_{\mbox{-}}^{\dagsup} X''-\W_{\mbox{-}}^{\dagsup}T_1^* +\a_0 +\a')$, 
where 
$K^{\dagsup}_{\mbox{-}}= K- \Kd$, $ K^{\ddagsup}_{\mbox{-}}= K+ \Kd$, and
$\W^{j}_{\mbox{-}}=K^j_{\mbox{-}}c$ are the Doppler shifted wavevectors and angular frequencies; 
$\a' =(2N+1)\pi$ as earlier.
Here at $X''$ and $T_1^*$, ${\varphi^{\dagsup}}^{\imc}_r$ meets the wave the virtual charge just generated to the left,   
${\varphi^{\ddagsup}}^{\imc}(X'',T_1^*)
=-C_1 \sin(K_{\mbox{-}}^{\ddagsup} X''+\W_{\mbox{-}}^{\dagsup}T_1^* -\a_0)$, and superpose with it to ${\widetilde \psi}^{\imc } (X,T_1^*)= {\varphi^{\dagsup}}^{\imc}_r +{\varphi^{\ddagsup}}^{\imc}$
$=2C_1\cos (KX''+K_d cT_1) \sin[-K_d X'' - 2\W T_1 -\a_0]$. 

Having $J>>1$ and being nondamping, $ {\widetilde \psi}^{\imc }$  
will be looping continuously, up to the present time $T$. 
Its present form ${\widetilde \psi}^{\imc }(X'',T)$ is then as if just produced by the virtual charge at time $T$ but at a location of a distance $L$ advancing the actual charge; it accordingly has a  phase advance   $\beta =\frac{(K^{\dagsup}-K^{\dagsup}_{\mbox{-}})}{2} L=K_d L $ relative to $\widetilde{\psi}$ (the phase advance in time yields no newer feature).
Including this $\beta$, using TI and with some algebra, 
${\widetilde \psi}^{\imc }(X'',T)$ writes on axis $X$ 
as  
\begin{eqnarray}\label{eq-psivir-W}
&{\widetilde \psi}^{\imc } (X,T)
=\widetilde{\Phimit}^{\imc}(X,T) \widetilde{\Psimit}^{\imc}(X,T),  
\\
& 
\widetilde{\Phimit}^{\imc}(X,T)  =  2C_1 \cos[(KX+2K_dcT ], 
\end{eqnarray}
\vspace{-0.6cm} 
$$\refstepcounter{equation} \label{eq-psivir-WPsi}
\widetilde{\Psimit}^{\imc} (X,T) = - \sin[\Kd X + (\W +\W_d)T +\a_0+\beta]. \eqno(\ref{eq-psivir-WPsi})
$$ 
   ${\widetilde \psi}^{\imc }$ of the virtual 
or reflected charge is seen to be similarly a traveling beat or de Broglie phase  wave to the left of a phase velocity $-W$, 
and of 
 $K_\beat$, ${\mit\Lambda}_\beat$ and $\W_\beat$ as of (\ref{eq-Wbeat1}).
\begin{figure}[h] 
\vspace{0.cm}
\begin{center}
\includegraphics[width=0.85\textwidth]{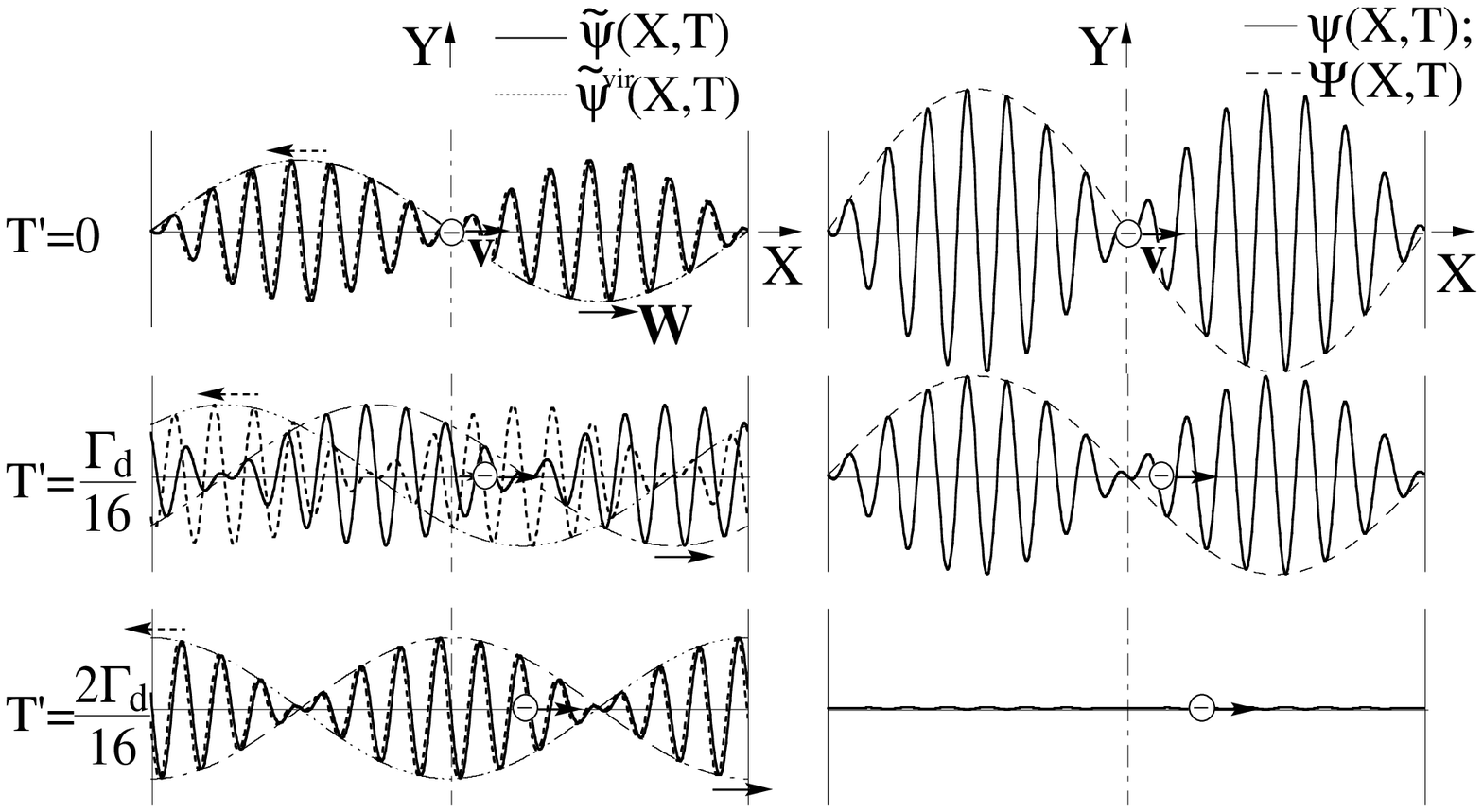}
\end{center}
\vspace{-1.05cm}
\begin{center}
\includegraphics[width=0.85\textwidth]{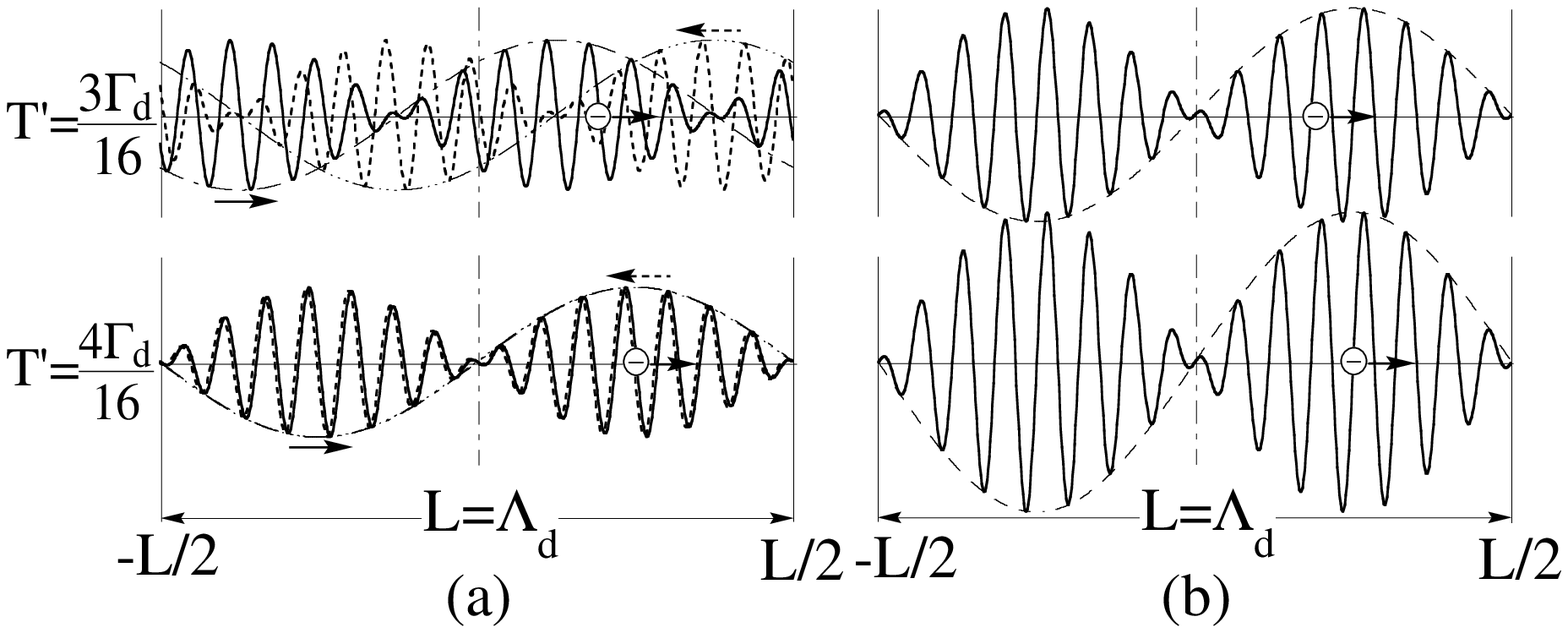}
\end{center}
\vspace{-0.5cm}
 \caption{ 
(a) The beat waves  $\widetilde{\psi}$ traveling at a phase velocity $W$ to the right as in Figure \ref{figI.4-dBwav-trv}b  and  $\widetilde{\psi}^{\imc}$ at $-W$ to the left, of a wavelength ${\mit\Lambda}_d$, due to the right- and left- traveling actual and virtual sources respectively.
(b)  $\widetilde{\psi}$ and  $\widetilde{\psi}^{\imc}$ superpose to a standing beat or de Broglie phase wave $\psi$ of wavelength ${\mit\Lambda}_d$, angular frequency $\sim\W$. Along with the $\psi$ process, the particle's center of mass ($\ominus$) is transported  at the velocity $v$, of a period $\frac{2\pi}{\W_d}={\mit\Lambda}_d/v$.
}
\label{fig-wvpac}
\vspace{-0.8cm}
\end{figure}

\section{Standing total wave and de Broglie wave}
Now if $K_d L(=\beta)=n \pi$, i.e.   
\begin{eqnarray}\label{eq-Kdn}
\label{eq-kd3}
\Kd_n = \frac{n\pi}{L}, \quad     n=1,2,\ldots, 
\end{eqnarray}
and accordingly 
 ${\mit\Lambda}_{dn}
=\frac{2L}{n}$, 
then  $ {\widetilde \psi}^{\imc }$ and  $ {\widetilde \psi}$ superposed onto themselves from different loops are each a maximum.   
Also, at ($X$, $T$), $ {\widetilde \psi}^{\imc }$ and  $ {\widetilde \psi}$ meet  and superpose, 
as $\psi
=\widetilde{\Phimit}\widetilde{\Psimit}+\widetilde{\Phimit}^{\imc}\widetilde{\Psimit}^{\imc}$. On the scale of ${\mit\Lambda}_d$, or $K_d$, the time variations in $\widetilde{\Phimit}$ and $\widetilde{\Phimit}^{\imc}$
are  higher-order ones; thus for $K>>K_d$, we have to a good approximation $ \widetilde{\Phimit}(X,T)\simeq
 \widetilde{\Phimit}^{\imc}(X,T)\simeq
 2C_1\cos (K X)=F(X)$. 
Thus   
$\psi (X,T)= F(X) [\widetilde{\Psimit}+\widetilde{\Psimit}^{\imc}]
= C_4\cos(KX)\sin[(\W +\W_d)T]   \cos (K_d X+\a_0)$; 
$C_4=4C_1$. 
The mechanical condition at the massive walls 
$\psi(0,T)=\psi(L,T)=0$
requires 
$\a_0=-\frac{\pi}{2}$. Hence finally  
\begin{eqnarray}\label{eq-beatstd}
&\psi(X,T)= \Phimit(X,T)\Psimit_\Xssub(X);  
\\ 
\label{eq-PsiA} 
&\Psimit_\Xssub(X)=\sin (K_d X),
\\  \label{eq-Phi}\label{eq-PhiA}
  & \Phimit(X,T)=C_4\cos(KX)\sin[(\W +\W_d)T].  
\end{eqnarray} 
$\psi$ of (\ref{eq-beatstd}) is a {\it standing beat}, or {\it standing de Broglie phase wave}; it includes all of the  component waves due to both the actual and virtual charges and hence represents the (standing) total  wave of the particle.

\section{Eigen-state wave function and variables}
\label{SecI.5.2.b} \label{SecI.Vb} \label{SecI.VIb}  
\label{Sec-NdBstnd}

\begin{figure}[t]
\vspace{0.2cm}
\begin{center}
\includegraphics[width=0.88\textwidth]{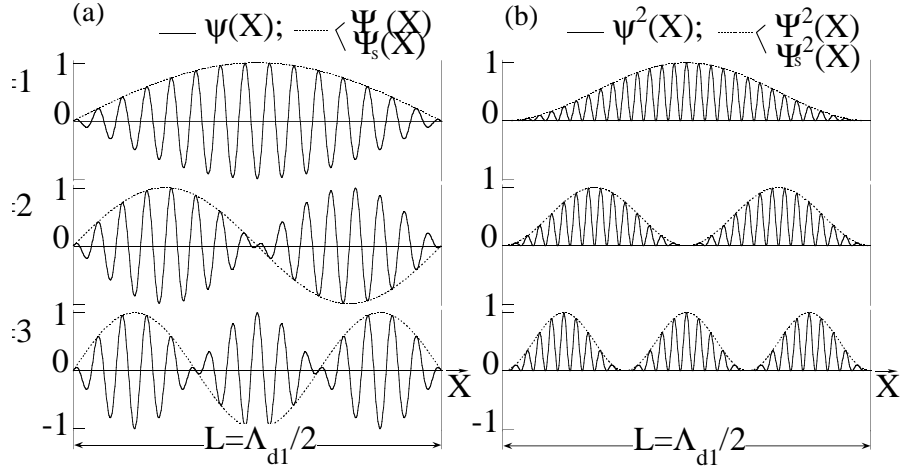}
\end{center}
\vspace{-0.7cm}
 \caption{
 (a)  The total wave of particle $\psi(X)$ of (\ref{eq-beatstd}) with rapid oscillation, 
and the de Broglie wave $\Psimit(X)$ as the envelop,  
for three lowest energy levels $n=1,2,3$; 
$\Psimit$ coincides with Schr\"odinger eigen-state functions $\Psimit_S$. (b) The corresponding probabilities.  
}\label{figI.6-dBwav-n}
\leavevmode\vspace{-0.5cm}
\end{figure}

Equation (\ref{eq-e3}) showed the particle's kinetic energy is transported at the angular frequency $\frac{1}{2}\W_d$, half the value $\W_d$ for transporting the particle mass, and is a source motion effect of order $(\frac{v}{c})^2$. This is distinct from, actually  exclusive of, the source motion effect, of order $v$, responsible for the earlier beat wave formation. We here include the order $(\frac{v}{c})^2$ effect only simply  as a multiplication factor to $\psi$,  and thus have $\psi' =\psi(X,T) e^{-i \frac{1}{2}\hbar \W_d T}$ which describes the particle's kinetic energy transportation. Furthermore, in typical applications  $K\gg \Kd$, $\W\gg \W_d$; thus on the scale of ($K_d$, $\W_d$), we can to a good approximation ignore the rapid oscillation in $\Phimit$ of (\ref{eq-PhiA}), and have 
$$
 \Phimit (X,T)
\simeq C_4 
\equiv{\rm Constant}   \eqno(\ref{eq-PhiA})'
$$ 
and $\psi (X,T)=C \Psimit_\Xssub(X)$. 
The time-dependent wave function, in energy terms,  is thus $\Psimit(X,T)$ $= \psi'(X,T)=\psi e^{-i\frac{\W_d}{2}T}=C \Psimit_\Xssub(X) e^{-i\frac{\W_d}{2}T}$, or
\begin{eqnarray}\label{eq-psipac}
 \Psimit(X,T)
         =C \sin (\Kd X)  e ^{-i \frac{1}{2} \W_d T},
\end{eqnarray}
where $C=\frac{1}{\int^L_0 \psi^2 d X}
=\frac{      \sqrt{2/L} }{ C_4}$ is a normalization constant.
With (\ref{eq-kd3}) for $K_{dn}$ in 
(\ref{eq-px3})--(\ref{eq-e3}),  
for a fixed $L$ we have the permitted dynamic variables 
\refstepcounter{equation}
 \label{eq-Pvn}
\refstepcounter{equation}   \label{eq-Evn}
$$\displaylines{
 \hfill \qquad  \Pm_{\vel n}=   \frac{nh } {2L},   \hfill
\quad \ \ (\ref{eq-Pvn})
\quad \qquad \hfill\hfill
        \Eng_{\vel n}   =\frac{n^2 {h}^2  }{8 M L^2}, \qquad \hfill(\ref{eq-Evn}) \nonumber 
}$$
where  $n=1,2,\ldots.$. 
These dynamic variables are seen to be quantized, pronouncingly  for $L$ not much greater than ${\mit\Lambda}_d$, as the direct result of the standing wave solutions.  As shown for the three lowest energy levels in Figure 
\ref{figI.6-dBwav-n}a, 
the permitted $\Psimit(X,T_0)\equiv \Psimit(X)$, with $T_0$ a fixed point in time, describing the envelopes (dotted lines) for $\psi(X,T_0)\equiv \psi(X)$ (solid lines) which rapid  oscillation has no physical consequence to the particle dynamics, 
  are in complete agreement with the corresponding solution to
Schr\"odinger equation for an identical system, 
$\Psimit_S(X)$ (the same dotted lines).

The total wave of a particle, hence its total energy $\Eng(X)$, mass, size, all  extend in (real) space throughout the wave path.  A portion of the particle, hence the probability in finding the particle,  at position $X$ in space is proportional to $\Eng(X)$ stored in the infinitesimal volume at $X$, 
$ \Eng (X)=B\psi^2(X) \propto \psi^2(X)$ (Fig\ref{figI.6-dBwav-n}b), with $B$ a conversion constant \cite{Unif1}.

With (\ref{eq-Pvn}) in $\Delta \Pm_{\vel} = \Pm_{\vel.n+1} -\Pm_{\vel.n} $ we have $  \Delta \Pm_{\vel} 2L = h $,
 which  reproduces Heisenberg's uncertainty relation. 
It follows from the solution that the uncertainty in finding a particle in real space results from the particle is an extensive wave over $L$, and in momentum space from the standing wave solution where waves interfering destructively are cancelled and inaccessible to an external observer. 

\section{Concluding remarks}
We have seen that the total wave superposed from the electromagnetic component  waves generated by a traveling oscillatory vaculeon charge, which together make up our particle, has actually the requisite properties of a de Broglie wave. It exhibits in spatial coordinate the periodicity of the de Broglie wave, by the wavelength ${\mit\Lambda}_d$, 
facilitated by a beat or de Broglie phase wave traveling at a phase velocity $\sim c^2/v$, with the beat in the total wave resulting naturally from the source-motion resultant Doppler differentiation of the electromagnetic component waves. ${\mit\Lambda}_d$ conjoined with the particle's center-of-mass motion leads to a periodicity of the de Broglie particle wave on time axis, the angular frequency $\W_d$. The ${\mit\Lambda}_d$ and $\W_d$ obey the de Broglie relations. The particle's standing wave solutions in confined space  agree completely with the Schr\"odinger solutions for an identical system.

\end{document}